\documentclass{emulateapj}
\usepackage{float}
\usepackage{epsfig}

\begin{document}
\title{On the Reversal of SFR-density Relation at z=1: Insights from Simulations}
\author{Stephanie Tonnesen$^1$ and Renyue Cen$^2$}
\affil{Department of Astrophysics, Princeton University, Peyton Hall, Princeton, NJ, 08544}
\email{1 stonnes@astro.princeton.edu\\
2 cen@astro.princeton.edu}

\begin{abstract}
Recent large surveys have found a reversal of the star formation rate (SFR)-density relation at $z$=1 from that at $z$=0 (e.g. Elbaz et al.; Cooper et al.), while the sign of the slope of the color-density relation remains unchanged (e.g. Cucciati et al.; Quadri et al.).  We use state-of-the-art adaptive mesh refinement cosmological hydrodynamic simulations of a 21$\times$24$\times$20 $h^{-3}$ Mpc$^3$ region centered on a cluster to examine the SFR-density and color-density relations of galaxies at $z$=0 and $z$=1.  The local environmental density is defined by the dark matter mass in spheres of radius 1 $h^{-1}$ Mpc, and we probe two decades of environmental densities.  Our simulations produce a large increase of SFR with density at z=1, as in the observations of Elbaz et al.  We also find a significant evolution to z=0, where the SFR-density relation is much flatter.  The color-density relation in our simulations is consistent from $z$=1 to $z$=0, in agreement with observations.  We find that the increase in the median SFR with local density at $z$=1 is due to a growing population of star-forming galaxies in higher-density environments.  At $z$=0 and $z$=1 both the SFR and cold gas mass are tightly correlated with the galaxy halo mass, and therefore the correlation between median halo mass and local density is an important cause of the SFR-density relation at both redshifts.  We also show that the local density on 1 $h^{-1}$ Mpc scales affects galaxy SFRs as much as halo mass at $z$=0.  Finally, we find indications that the role of the 1 $h^{-1}$ Mpc scale environment reverses from $z$=0 to $z$=1:  at $z$=0 high-density environments depress galaxy SFRs, while at $z$=1 high-density environments tend to increase SFRs.
\end{abstract}
\section{Introduction}

In the nearby universe, the morphology, color, star formation rate (SFR), and mass of galaxies is correlated with local environmental density.  Specifically, galaxies tend to be earlier types, more massive, redder and to have lower SFRs and specific SFRs (sSFR $\equiv$ SFR/M$_*$) in higher density environments (Dressler 1980; Oemler 1974; Balogh et al. 2001; Blanton et al. 2003; Hogg et al. 2004; Balogh et al. 1998; Hashimoto et al. 1998; G{\'o}mez et al. 2003).  While galaxy morphology, color, and SFR also depend on mass (e.g. de Vaucouleurs 1960; de Vaucouleurs 1962; Strateva et al. 2001; Blanton et al. 2003; Baldry et al. 2004), Kauffmann et al. (2004) bins galaxies by mass and finds that star formation history (SFH) depends on the local density at any galaxy stellar mass.

The first evidence that the galaxy population in clusters may evolve with redshift was found by Butcher \& Oemler (1978), who observed that galaxies in clusters beyond $z$$\sim$0.4 tend to be bluer than galaxies in nearby clusters.  However, the Butcher \& Oemler observations did not compare the cluster to the field populations at high redshift, while later observations showed that galaxies in all environments had higher SFRs in the past (Lilly et al. 1996; Madau et al. 1996; Wilson et al. 2002).  Using large surveys able to probe a range of environments at high redshifts, recent observations have found that the SFR-density and color-density relations may be reversed at $z$$\sim$1 compared to those at $z$$\sim$0 (e.g. Cucciati et al. 2006; Elbaz et al. 2007; Cooper et al. 2008; Ideue et al. 2009; Salimbeni et al. 2009; Tran et al. 2010; Gr{\"u}tzbauch et al. 2011; Popesso et al. 2011).  

For example, Elbaz et al. (2007; E07) compare local star-forming SDSS galaxies to star-forming galaxies in the GOODS-S and GOODS-N fields at $z$$\sim$ 1.  They find a reversal of the SFR-density relation: specifically that the median $<$SFR$>$ decreases with increasing galaxy density in the local sample, while at $z$$\sim$1, the median $<$SFR$>$ first  increases with increasing galaxy density before peaking and then decreasing again at the highest densities.  They claim that the increase of $<$SFR$>$ with local galaxy density is caused by increases of both M$_*$ and sSFR with galaxy density.  Because in general sSFR decreases with increasing M$_*$ (e.g. Blanton et al. 2003; Baldry et al. 2004), they conclude that at $z$$\sim$1 the environment must enhance the SFR of galaxies in order for sSFR to increase while M$_*$ increases.

Cooper et al. (2008; C08) also observed the reversal of the SFR-density relation by comparing a local SDSS sample to the $z$$\sim$1 DEEP2 sample (not limited to star-forming galaxies), but came to a somewhat different conclusion.  Because they find that the sSFR decreases with local density at both $z$$\sim$0 and $z$$\sim$1, they conclude that the reversal of the SFR-density relation is mainly due to the different luminosity-density (and mass-density) relations of the blue galaxies in their samples at the two redshifts:  at $z$$\sim$1 M$_{\mathrm{B}}$ increases with local density, while at $z$$\sim$0, M$_{\mathrm{B}}$ is constant with local density.

Several observations of individual clusters at $z$$\sim$1.5 find that they already have a significant passive galaxy population (e.g. McCarthy et al. 2007; Kurk et al. 2009; Wilson et al. 2009; Strazzullo et al. 2010), indicating that at least the color-density relation extends beyond $z$$\sim$1.  Further, using an early release of the UKIDSS Ultra-Deep Survey, Chuter et al. (2011) find a strong relationship between rest-frame (U $-$ B) color and galaxy environment to $z$$\sim$1.5, with red galaxies residing in significantly denser environments than blue galaxies on scales below 1 Mpc.  Using a more complete UKIDSS Ultra-Deep Survey data release, Quadri et al. (2012; Q12) use a mass-selected sample to show that galaxies with quenched SF tend to reside in dense environments out to as least $z$$\sim$1.8.  Specifically, they find that the quiescent fraction increases with local density, even at high redshift.  

It is difficult to synthesize these observational results into a single coherent picture, as they use different wavelengths to measure color and/or SFR, use different methods to measure local density, observe different fields and define their galaxy samples differently.  In addition, semi-analytic models (SAMs) have not reproduced observations of the reversal of the SFR-density or color-density relations (E07; Cucciati et al. 2006; 2012).    

In this paper we present the SFR-density, sSFR-density, and quiescent fraction of galaxies at $z$ = 0 and  $z$ $=$ 1 in a hydrodynamic cosmological simulation.  We choose these redshifts in order to compare our results to recent work, specifically E07, C08, and Q12. 
We focus on one region centered around a cluster within a large simulation with box side of 120 $h^{-1}$ Mpc, with size 21$\times$ 24$\times$20 $h^{-3}$ Mpc$^3$, which gives us a broad range of local environments to study.  Our goal is to gain physical insight into what sets galaxy SFRs by comparing the SFR, sSFR, and quiescent fraction as a function of environmental density on the same galaxy sample.  

After a brief description of  our simulations (Section \ref{sec:boxes}), we discuss our galaxy selection technique and our method for determining quantities for each galaxy in Section \ref{sec:galaxyselection}.  In Section \ref{sec:localdensity} we define our local density parameter.  In Section \ref{sec:obscomparisons} we compare our results to E07 (Section \ref{sec:E07}), C08 (Section \ref{sec:C08}) and Q12 (Section \ref{sec:Q12}).  We then discuss and interpret our results in terms of the halo mass (\S \ref{sec:story}), and highlight why our results differ from those based on semi-analytic models in Section \ref{sec:simcomparison}.  Finally, we summarize our main conclusions (\S \ref{sec:conclusion}).

\section{Methodology}
\subsection{The Simulations}\label{sec:boxes}

For the details of our simulations, we refer the reader to Cen (2012), although for completeness we reiterate the main points here.  We perform cosmological simulations with the adaptive mesh refinement (AMR) Eulerian hydrodynamical code \textit{Enzo} (Bryan 1999; O'Shea et al. 2004; Joung et al. 2009).  We use cosmological parameters consistent with the WMAP7-normalized LCDM model (Komatsu et al. 2011):  $\Omega_M$ = 0.28, $\Omega_b$ = 0.046, $\Omega_\Lambda$ = 0.72, $\sigma_8$ = 0.82, $H_o$ = 100 $h$ km s$^{-1}$ Mpc$^{-1}$ = 70 km s$^{-1}$ Mpc$^{-1}$, and $n$ = 0.96.  We first ran a low resolution simulation with a periodic box of 120 $h^{-1}$ Mpc on a side, and identified a region centered on a cluster at $z = 0$.  We then resimulated the region with high resolution, but embedded within the outer 120 $h^{-1}$ Mpc box to properly take into account large-scale tidal field effects and appropriate fluxes of matter, energy and momentum across the boundaries of the refined region.

The refined region we discuss in this paper is 21$\times$24$\times$20 $h^{-3}$ Mpc$^3$ centered on a cluster of $\sim$2$\times$10$^{14}$ M$_\odot$ with a virial radius (r$_{200}$) of 1.3 $h^{-1}$ Mpc.  We emphasize that this highly-resolved box is much larger than the cluster at its center, and that therefore there are galaxies in a range of local densities.  In Tonnesen \& Cen (2012) we compare the local densities of galaxies in this refined region to a different refined region in the same large periodic box centered around a void, and showed that there is substantial overlap in the distribution of local densities found in the two very different large-scale environments.

We consider two refined simulations of the same region that differ in their level of maximum refinement.  In the refined region in the low resolution simulation, the minimum cell size is 0.46 $h^{-1}$ kpc, using 11 refinement levels at $z = 0$.  The initial conditions for the refined region include a mean interparticle separation of 117 $h^{-1}$ kpc comoving and a dark matter particle mass of 1.07$\times$10$^8$ $h^{-1}$ M$_\odot$.  In the high resolution simulation, the minimum cell size is 0.114 $h^{-1}$ kpc, using 13 refinement levels at $z = 0$, with an initial mean interparticle separation of 58.6 $h^{-1}$ kpc comoving and a dark matter particle mass of 1.5$\times$10$^7$ $h^{-1}$ M$_\odot$.  We only have the $z$$=$1 output of the HR simulation, because it has not yet run to $z$$=$0.   

The simulations include a metagalactic UV background (Haardt \& Madau 1996), a model for shielding of UV radiation by neutral hydrogen (Cen et al. 2005), and metallicity-dependent radiative cooling (Cen et al. 1995).  The fraction and density of neutral hydrogen is directly computed within the simulations.  Star particles are created in gas cells that satisfy a set of criteria for star formation proposed by Cen \& Ostriker (1992), and reiterated with regards to this simulation in Cen (2012).  Once formed, the stellar particle loses mass through gas recycling from Type II supernovae feedback, and about 30\% of the stellar particle mass is returned to the ISM within a time step.  Supernovae feedback is implemented as described in Cen (2012):  feedback energy and ejected metal-enriched mass are distributed into 27 local gas cells centered at the star particle in question, weighted by the specific volume
of each cell, which is to mimic the physical process of  supernova blastwave propagation that tends to channel energy, momentum, and mass into the least dense regions (with the least resistance and cooling). We allow the whole feedback process to be hydrodynamically coupled to surroundings and subject to
relevant physical processes, such as cooling and heating, as in
nature.  The low resolution simulation used in this paper has compared several galaxy properties that depend critically on the feedback method to observations and found strong agreement (Cen 2011a-c; Cen 2012).  We do not include a prescription for AGN feedback in this simulation, and as a result, our simulation overproduces luminous galaxies at the centers of groups and clusters of galaxies.  This is discussed in detail in Cen (2011c), who shows that the luminosity function of the simulated galaxies agrees well with observations at $z$=0 except at the high-luminosity end.  When Cen (2011c) adds an AGN feedback correction in the post-simulation analysis for halos with masses greater than 10$^{13}$ M$_{\odot}$ or for galaxies with stellar masses above 4$\times$10$^{12}$ M$_{\odot}$, the simulated luminosity function also agrees with observations at the high luminosity end.  In the low resolution run, each star particle has a mass of $\sim$10$^6$ M$_\odot$, which is similar to the mass of a coeval globular cluster.  In the high resolution run, each star particle has a mass of $\sim$10$^5$ M$_\odot$.

\subsection{Galaxies}\label{sec:galaxyselection}

We use HOP (Eisenstein \& Hut 1998) to identify galaxies using the stellar particles.   HOP uses a two-step procedure to identify individual galaxies. First, the algorithm assigns a density to each star particle based on the distribution of the surrounding particles and then hops from a particle to its densest nearby neighbor until a maximum is reached. All particles (with densities above a minimum threshold, $\delta_{outer}$) that reach the same maximum are identified as one coherent group. In the second step, groups are combined if the density at the saddle point which connects them is greater than $\delta_{saddle}$). We use HOP because of its physical basis, although we expect similar results would be found using a friends-of-friends halo finder.  This has been tested and is robust using reasonable ranges of values (e.g. Tonnesen, Bryan, \& van Gorkom 2007).

After we identify galaxies using stellar particles, we create our sample using only well-resolved galaxies within our refinement region.  First, we restrict our sample to galaxies that have resided in the refined region since the beginning of the simulation.  To do this, we require that all of the dark matter particles within the virial radius (in detail, we use r$_{200}$, the radius within which the average density is 200 times the critical density, which is directly calculated using simulation output) of each galaxy be the low-mass dark matter particles with which we populated the refined region of our simulation.  Second, we only consider those galaxies with a stellar mass greater than 5$\times$10$^{9}$ M$_\odot$.  We choose this lower limit to our stellar mass because we have found that above this mass our sample is more than 75\% complete (Cen 2014).  We have tested that including lower mass galaxies down to 10$^8$ M$_{\odot}$ has little quantitative and no qualitative effect on our results.  Implementing these two criteria leaves 61\% of the originally identified galaxies at $z$=0 in the low resolution run. 

We plotted projections of the star particles of each of the galaxies identified by HOP at $z$=0 in order to verify first that HOP was identifying galaxy-like objects (with a stellar density peak), and second, that HOP was not grouping multiple galaxies together.  Only a few of the galaxies above our minimum mass that HOP identified had density profiles without a strong density peak, and a small number of HOP-identified galaxies in fact had two density peaks, and one had three peaks.  Both of these problems add up to a misidentification of only 2\%.  

In this paper we measure the stellar mass (M$_*$), dark matter halo mass (M$_{\mathrm{halo}}$), cold gas mass (M$_{\mathrm{cold}}$), SFR, M$_{\mathrm{B}}$, and $g$ $-$ $r$ color.  Stellar mass is determined by adding the mass of each star particle identified by HOP to belong to a galaxy.  The dark matter halo mass is calculated by summing the mass of all the dark matter particles out to r$_{200}$.  

The cold gas mass is defined slightly differently in the low resolution versus the high resolution simulation due to differences in how the galaxy data sets were originally made, although as we now discuss the differences are minor for the purposes of this paper.  In the low resolution simulation, M$_{\mathrm{cold}}$ is the sum of all HI gas within r$_{200}$.  As noted in Section \ref{sec:boxes}, the fraction and density of neutral hydrogen is directly computed within the simulations.  The high resolution data set includes values of the total gas within 100 kpc and 300 kpc with a temperature less than 10$^5$ K.  Because this gas is at the peak of the cooling curve it quickly cools to $\sim$10$^4$ K, and there is therefore little effective difference in the temperatures of the two cold gas definitions.  However, rather than use one or the other of these set radii, in order to compare with the M$_{\mathrm{cold}}$ from the low resolution run we want to include all of the cold gas within r$_{200}$.  Therefore, in order to estimate the cold gas within the virial radius of galaxies in the high resolution simulation, we separate the galaxies into low-mass galaxies with M$_{\mathrm{halo}}$ $<$ 2.5$\times$10$^{11}$M$_{\odot}$ (and therefore virial radii at or less than 100 kpc), high-mass galaxies with M$_{\mathrm{halo}}$ $>$ 6$\times$10$^{12}$M$_{\odot}$ (and virial radii at or greater than 300 kpc), and galaxies that fall between these two extremes.  The two extreme cases are simple, and M$_{\mathrm{cold}}$ is either the gas within 100 kpc for the low-mass galaxies or within 300 kpc for the high-mass galaxies.  For all other galaxies, we use the halo mass to determine the fraction of gas between 100-300 kpc that we will include in M$_{\mathrm{cold}}$, using M$_{\mathrm{cold}}$ = M$_{100 kpc}$ + (M$_{\mathrm{halo}}$/6$\times$10$^{12}$)$\times$(M$_{300 kpc}$-M$_{100 kpc}$).  With these definitions,  M$_{\mathrm{cold}}$ is consistent between the low- and high-resolution simulations at $z$=1.

The luminosity of each stellar particle at each of the Sloan Digital Sky Survey (SDSS) five bands is computed using the GISSEL stellar synthesis code (Bruzual \& Charlot 2003), by supplying the formation time, metallicity and stellar mass.  M$_{\mathrm{B}}$ is calculated as M$_{\mathrm{B}}$ $=$ $\mathrm{M_g}$ + 0.313$\times$$(\mathrm{M_g}-\mathrm{M_r}) + 0.2271$ (Lupton 2005).  The SFR for a galaxy is calculated based on the creation time and mass of each star particle.
  
\subsection{Local Density}\label{sec:localdensity}

We choose to use dark matter mass to define the local environment because it is a fundamental measure of the local mass in a region.  Further, using dark matter mass is not plagued by the uncertainties in using galaxy counts as a measure of local density. In simulations the galaxy number count could be affected by resolution issues that may affect galaxy number counts, while in observational samples local galaxy counts may be affected by projection effects due to lack of means to distinguish between peculiar velocity and Hubble velocity.  

We define the local environment by measuring the dark matter mass in spheres with a comoving radius of 1 $h^{-1}$ Mpc.  In this paper we focus on how galaxy SFRs and gas content may be affected on scales larger than the virial radius (all halos except the cluster halo have virial radii less than 1 $h^{-1}$ Mpc), for example whether residing in collapsed filaments and pancakes would affect these properties (Cen 2011c).  This is done by summing the dark matter mass directly, not demanding that it is within the virial radius of any galaxy.  We randomly select 27 000 positions within the refined region (with a 2 $h^{-1}$ Mpc buffer at each edge) to be the center of the spheres.  Thus we oversample our box and end up with a total of 27 000 environments that we probe, which we differentiate by the dark matter mass in the sphere, M$_{\mathrm{sph}}$.  However, not every r=1 $h^{-1}$ Mpc sphere will contain a galaxy, so when we determine galaxy properties as a function of environment, we only consider those spheres that contain galaxies.  This technique of choosing random spheres in space rather than spheres centered on galaxies differs from that used by many observers, however, E07 compare average SFRs found in regions centered on galaxies to those found in regions centered on grid points independent of galaxy positions, and find negligible difference between the two techniques.

\section{Results}\label{sec:obscomparisons}

\subsection{The reversal of the $<$SFR$>$-density relation}\label{sec:E07}

We first determine whether our SFR-density relation evolves from $z$$=$0 to $z$$=$1.  To facilitate comparisons to E07 we use similar galaxy selection criteria and methods.  E07 study the evolution of the SFR-density relation by comparing local SDSS galaxies to the GOODS-S and GOODS-N fields at $z$ $\sim$ 1.  They use a magnitude cut in both samples, M$_{\mathrm{B}}$ $\le$ -20, which they state is equivalent to a stellar mass cutoff of M$_*$ $\ge$  10$^{9}$ M$_{\odot}$ at $z$$=$0.8 and M$_*$ $\ge$ 10$^{10}$ M$_{\odot}$ at $z$$=$ 1.2 (the two extremes of their high-redshift range).  The local galaxy density is determined by counting all the galaxies within boxes centered on each galaxy of 1.5 comoving Mpc on a side, in a velocity interval $\Delta$$v$ $=$ $\Delta$$z$/(1 + $z$), $\Delta$$z$ $=$0.02.  E07 only include star-forming galaxies in their sample, down to SFR$_{UV}$ limits of 0.5 M$_{\odot}$ yr$^{-1}$ for the $z$$\sim$1 sample.  They use the star-forming galaxy sample and SFRs calculated in Brinchmann et al. (2004) for the low-redshift SDSS sample.  

E07 first calculates an average SFR ($<$SFR$>$) and local galaxy density in boxes centered on every galaxy.  Then they group the galaxies into local density bins and find the median of the distribution of $<$SFR$>$.  They find that the median $<$SFR$>$ decreases with increasing galaxy density in the SDSS sample, but, at $z$$\sim$1, the median $<$SFR$>$ first increases with increasing galaxy density before peaking and then decreasing at the highest local densities.  These results are not reproduced by a SAM using the Millennium simulation.  In the SAM, there is never a reversal of the SFR-density relation, just a gradual flattening (it is not flat by $z$$=$1).  When only examining galaxies with 5$\times$10$^{10}$ $<$ M$_*$/M$_{\odot}$ $<$ 5$\times$10$^{11}$, E07 finds that the sSFR increases with increasing local density.  They claim that the increase of $<$SFR$>$ with local density is due to increases of both M$_*$ and sSFR with galaxy density.  Because in general sSFR decreases with increasing M$_*$ (e.g. Blanton et al. 2003; Baldry et al. 2004), they find it likely that the environment enhances the SFR of massive galaxies.

In order to examine comparable galaxy samples, we examine galaxies from our $z$=1 and $z$=0 outputs and use the same magnitude cut as E07 (M$_{\mathrm{B}}$ $\le$ -20).  As discussed in Section \ref{sec:galaxyselection}, we also only include galaxies with M$_*$ $>$ 5$\times$10$^{9}$ M$_{\odot}$.  All galaxies that fulfill our magnitude and mass criteria are included in our sample.  Recall that the M$_{\mathrm{B}}$ cut in E07 is equivalent to an M$_*$ cutoff of 10$^9$ M$_\odot$ and 10$^{10}$ M$_\odot$ at $z$$=$0.8 and $z$$=$1.2, respectively, so the lower mass cut is well-matched to the GOODS-S and GOODS-N samples.  We separate our galaxies into local density (M$_{\mathrm{sph}}$) bins as defined in Section \ref{sec:localdensity}.

\begin{figure}
\includegraphics[trim= 0mm 0mm 0mm 0mm, clip]{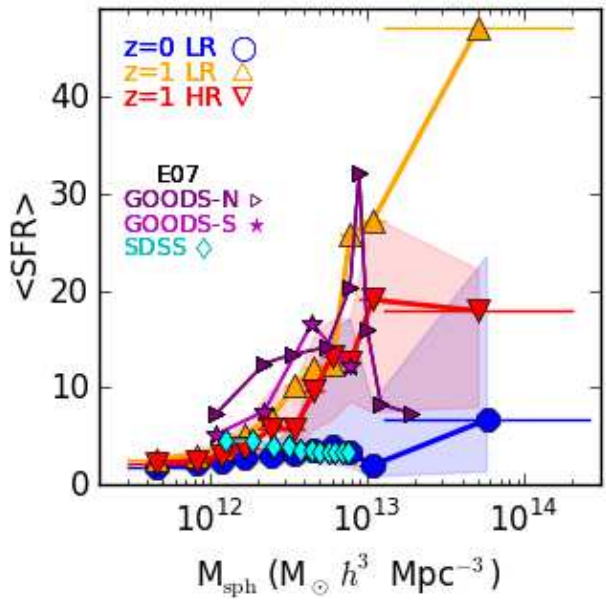}\\%
\caption{The median values of the $<$SFR$>$ in bins defined by dark matter mass (M$_{\mathrm{sph}}$) as a measure of local density.  The shaded regions denote the 25th-75th percentiles of the $<$SFR$>$ distribution at $z$=0 LR (blue) and $z$=1 HR (red).  This figure plots galaxies with M$_{\mathrm{B}}$ $\le$ -20, and 5$\times$10$^{9}$ $<$ M$_*$/M$_{\odot}$.  We overplot values from the E07 paper--in magenta and purple we plot the GOODS-S and GOODS-N samples, respectively, and in cyan we plot the SDSS sample.  We define the highest local density measurement in the GOODS-N bin to be in a DM halo of $\sim$1.5$\times$10$^{13}$ M$_{\odot}$, and assume that $\Sigma_{gal}$ $\propto$ M$_{\mathrm{sph}}$.} \label{fig-elbaz}
\end{figure}

We then plot $<$SFR$>$ as in E07, first calculating the average SFR ($<$SFR$>$) in each r$=$1 $h^{-1}$ Mpc sphere, then binning the spheres by local density (M$_{\mathrm{sph}}$) so that each bin has at least 100 spheres in each simulation output, and finally finding the median $<$SFR$>$ in each local density bin.  Results are shown as blue $\bigcirc$ ($z$=0 low resolution (LR)), orange $\bigtriangleup$ ($z$=1 LR), and red $\nabla$ ($z$=1 high resolution (HR)) in Figure \ref{fig-elbaz}.  The horizontal bars denote the M$_{\mathrm{sph}}$ bin sizes.  The shaded regions enclose the 25th to the 75th percentiles of the $z$=1 HR (red) and $z$=0 LR (blue) samples, denoting the (lower and upper) limits of the upper and lower quartiles of the data.    

We find that in the $z$=1 HR output, the $<$SFR$>$ increases with increasing local density from M$_{\mathrm{sph}}$$\sim$5$\times$10$^{11}$ M$_{\odot}$ to M$_{\mathrm{sph}}$$\sim$10$^{13}$ M$_{\odot}$, then decreases towards the highest M$_{\mathrm{sph}}$.  In the $z$=1 LR output, the $<$SFR$>$ increases up to the highest M$_{\mathrm{sph}}$.  At $z$=0 the $<$SFR$>$-density relation is nearly flat until the highest M$_{\mathrm{sph}}$ bin, at which point the $<$SFR$>$ increases.  We compare the $<$SFR$>$ in the lowest and highest M$_{\mathrm{sph}}$ bins, and find that the comparison is quite different in the $z$$=$1 and $z$$=$0 samples.  At $z$=1 the $<$SFR$>$ in the highest M$_{\mathrm{sph}}$ bin is nearly an order of magnitude more than the $<$SFR$>$ in the lowest M$_{\mathrm{sph}}$ bin, while at $z$=0 the $<$SFR$>$ in the highest M$_{\mathrm{sph}}$ bin is three times that in the lowest M$_{\mathrm{sph}}$ bin.  

In Figure \ref{fig-elbaz} we include all galaxies that fulfill our luminosity and mass criteria, including both star-forming and passive galaxies.  However, our result that the $<$SFR$>$-density relation is much steeper at $z$=1 than at $z$=0 is robust to whether or not we include only star-forming galaxies.  If we limit our sample to galaxies with SFR $>$ 0.5 M$_{\odot}$ yr$^{-1}$ (at both $z$=0 and $z$=1), the median $<$SFR$>$ is unchanged up to M$_{\mathrm{sph}}$ $=$ 3$\times$10$^{12}$ M$_{\odot}$, and the median $<$SFR$>$ in our five highest density bins increases by at most 33\% in the $z$=1 (HR or LR) outputs, and by as much as a factor of two in our $z$=0 output.  The $z$=1 $<$SFR$>$ remains universally higher than the $z$=0 $<$SFR$>$.

For comparison, we also plot values from E07 Figure 8.  The qualitative agreement between the trends at $z$=1 is clear, although there is a large amount of uncertainty in the alignment and stretch in the x-axis.  In order to compare our density measure (M$_{\mathrm{sph}}$) directly with that used in E07, we re-scale the E07 density values.  The largest group in the E07 $z$=0.8-1.2 sample has L$_{\mathrm{X}}$[0.5-2 keV] = 2$\times$10$^{42}$ erg s$^{-1}$, so is about 2-3$\times$10$^{13}$ M$_{\odot}$ (Stanek et al. 2006).  Thus, we choose the highest local density bin in the GOODS-N sample to be in a M$_{\mathrm{sph}}$ bin centered at 1.5$\times$10$^{13}$ M$_{\odot}$, and assume that there is a one-to-one relationship between the 2D ($\Sigma_{gal}$ in E07) and 3D (M$_{\mathrm{sph}}$ in this work) local density measures--i.e. $\Sigma$ $\propto$ $\rho$.

\begin{figure}
\includegraphics[trim= 0mm 0mm 0mm 0mm, clip]{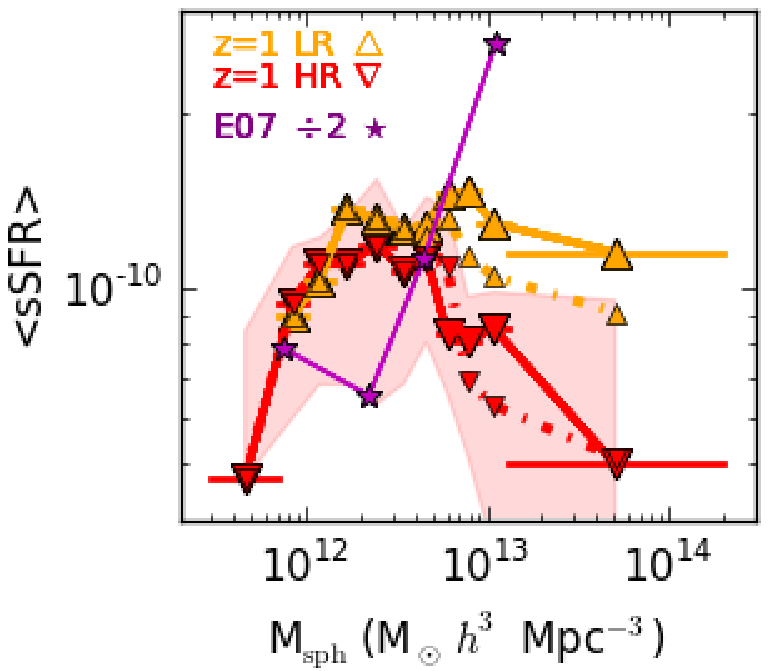}\\%
\includegraphics[trim= 0mm 0mm 0mm 0mm, clip]{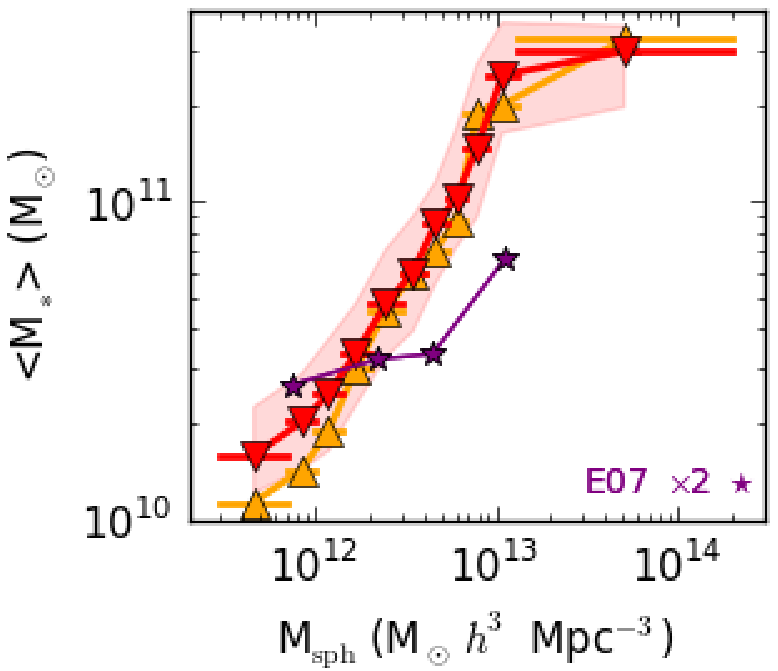}%
\caption{\textbf{Top:}  The median and mean values of the $<$sSFR$>$ in bins defined by dark matter mass (M$_{\mathrm{sph}}$) as a measure of local density.  This panel plots galaxies with M$_{\mathrm{B}}$ $\le$ -20, and 5$\times$10$^{10}$ $<$ M$_*$/M$_{\odot}$, as in the bottom panel of Figure 20 from E07 (the E07 values are overplotted).  To show that this result is not dominated by massive galaxies, we overplot galaxies with 5$\times$10$^{10}$ $<$ M$_*$/M$_{\odot}$ $<$ 5$\times$10$^{11}$ as dash-dotted lines.  The $<$sSFR$>$ increases, then decreases. 
\\ \textbf{Bottom:}  The median values of the $<$M$_*$$>$ in M$_{\mathrm{sph}}$ bins.  As in E07, we include galaxies with M$_{\mathrm{B}}$ $\le$ -20, and 5$\times$10$^{9}$ $<$ M$_*$/M$_{\odot}$, and find that the mass of galaxies increases with local density (M$_{\mathrm{sph}}$).
}\label{fig-essfr}
\end{figure}

In agreement with observed local mass-density relations (e.g. Balogh et al. 2001; Blanton et al. 2003), E07 find that the average stellar mass of 0.8$<$$z$$<$1.2 galaxies increases as a function of local density (top panel of their Figure 20).  Based on the relationship between mass and sSFR (e.g. Blanton et al. 2003; Baldry et al. 2004), this indicates that the sSFR of galaxies should \textit{decrease} as a function of increasing local density.  However, when they consider only massive galaxies at $z$$=$1, specifically 5$\times$10$^{10}$ $<$ M$_*$/M$_{\odot}$ $<$ 5$\times$10$^{11}$, the average sSFR increases as a function of local density (bottom panel of their Figure 20).  Thus, E07 conclude that the local environment causes the increase in the $<$SFR$>$ at higher local densities at $z$=1.

In order to determine if our results conform to the interpretation in E07, in the top panel of Figure \ref{fig-essfr} we plot the median of the $<$sSFR$>$ of the massive galaxies in our sample (5$\times$10$^{10}$ $<$ M$_*$/M$_{\odot}$) binned by local density (M$_{\mathrm{sph}}$).  Because our lack of AGN feedback will increase the SFR in massive galaxies, thus causing an increasing $<$sSFR$>$-density relation, we also plot the $<$sSFR$>$-density relation for 5$\times$10$^{10}$ $<$ M$_*$/M$_{\odot}$ $<$ 5$\times$10$^{11}$ galaxies.  We choose this mass range because it matches the masses studied in E07 and because the SFRs we measure would not be lowered by AGN feedback.  The lower mass range is shown as the dashed lines connecting the smaller symbols.  We also overplot values from Figure 20 in E07, scaled by a factor of two.  We scale the values so that the relative shapes from the observations and simulations can be easily compared.  

In the bottom panel of Figure \ref{fig-essfr} we plot the median $<$M$_*$$>$ of the 5$\times$10$^{9}$ $<$ M$_*$/M$_{\odot}$ total sample, which also increases with increasing M$_{\mathrm{sph}}$ (the median $<$M$_*$$>$ of the 5$\times$10$^{10}$ $<$ M$_*$/M$_{\odot}$ sample also increases with increasing M$_{\mathrm{sph}}$).  We overplot the M$_*$ values from E07, multiplied by two.  Because our simulated $<$SFR$>$ values are similar to those in observations (cf. Figure \ref{fig-elbaz}), our lower $<$sSFR$>$ values indicate that our simulated stellar masses are too high.  Indeed, we find better agreement between the simulated and observed stellar masses when we multiply the observed M$_*$ by two.  

Our $<$sSFR$>$ increases by about a factor of two while the $<$sSFR$>$ in E07 increases by a factor of about four.  Further, the shape of our sSFR-density relation is quite different from that in E07 as it is flat from 10$^{12}$$<$M$_{\mathrm{sph}}$/M$_{\odot}$$<$5$\times$10$^{12}$.  However, as in E07, we find that although M$_*$ increases, sSFR does not decrease up to at least M$_{\mathrm{sph}}$$\sim$5$\times$10$^{12}$.  This is true for either mass range or resolution we consider.  Since we expect an increase in M$_*$ to drive a decrease in sSFR, our $<$sSFR$>$-density relation may be flatter than that in E07 because our $<$M$_*$$>$-density relation increases more steeply (Figure \ref{fig-essfr}).  We find that our stellar masses tend to be larger and increase more quickly with increasing local density.  We speculate that rerunning our simulation with a lower star formation efficiency might lower early SFRs, producing lower-mass galaxies.  

\subsection{Comparing the SFR-density and sSFR-density relations}\label{sec:C08}

\begin{figure}
\includegraphics[trim= 0mm 0mm 0mm 0mm, clip]{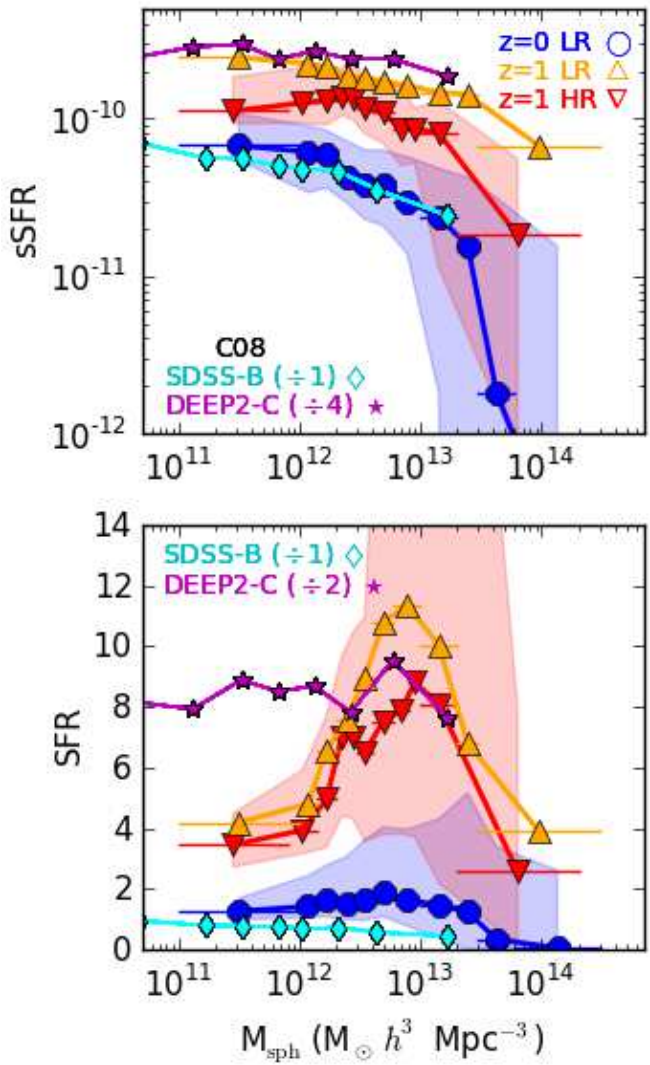}%
\caption{The median values (with the shaded area denoting the 25th-75th percentiles) of the sSFR and SFR in local density bins defined by dark matter mass.  This figure uses a redshift-dependent magnitude cut that is M$_{\mathrm{B}}$ $\le$ -20.53 at $z$=1 and M$_{\mathrm{B}}$ $\le$ -19.16 at $z$=0, following C08.  We again only include galaxies with 5$\times$10$^{9}$ $<$ M$_*$/M$_{\odot}$.  We overplot estimated values from C08--the DEEP2-C sample in purple, and the SDSS-B sample in cyan.  In order to compare the shapes of the relations, the sSFR in the DEEP2-C sample is scaled by a factor of 4 (divided by 4), and the SFR in the DEEP2-C sample is scaled by a factor of 3.  We define the highest local density measurement in DEEP2 to be in a DM sphere of 2$\times$10$^{13}$ M$_{\odot}$, and assume that $\Sigma_{gal}$ $\propto$ M$_{\mathrm{sph}}$.  
}\label{fig-Cooper}
\end{figure}
We next use a galaxy sample matched to that in C08 to compare the SFR-density relation to the sSFR-density relation.  As in C08, in this section at each redshift we use the same galaxy sample when we measure the SFR and sSFR as a function of density.  C08 used SDSS galaxies at low redshift and the DEEP2 sample at $z$$\sim$1.  The color-independent completeness limit of the DEEP2 survey at $z$ = 1.05 is M$_{\mathrm{B}}$ $\le$ -20.6, so in their DEEP2-C sample they use a redshift-dependent magnitude cut that is M$_{\mathrm{B}}$ $\le$ -20.53 at $z$=1 and M$_{\mathrm{B}}$ $\le$ -19.16 at $z$=0 (SDSS-B sample).  We choose to compare to this magnitude-limited observed sample because the magnitude limit acts to remove the lowest-mass galaxies from their sample (see C08 Figure 7).  This results in a closer match to the mass range in our simulated sample, which, as discussed in Section \ref{sec:galaxyselection}, has a minimum stellar mass of 5$\times$10$^{9}$ M$_{\odot}$. 

Our median sSFR and SFR as a function of local density (M$_{\mathrm{sph}}$) are shown in Figure \ref{fig-Cooper}.  As in C08, we simply take the median SFR or sSFR of all galaxies that fall into a bin of local density (M$_{\mathrm{sph}}$) (rather than the median of the average SFR or sSFR in each M$_{\mathrm{sph}}$ as in Section \ref{sec:E07}).  As in the previous figures, the $z$=0 LR is blue $\bigcirc$, $z$=1 LR is orange $\bigtriangleup$, and $z$=1 HR is red $\nabla$.  The horizontal lines denote the M$_{\mathrm{sph}}$ bin sizes.  The shaded regions enclose the region between the 25th and 75th percentiles of the $z$=1 HR (red) and $z$=0 LR (blue) samples.  Our bins are selected to include at least 100 galaxies.  We overplot estimates of the values from Figures A1 \& A2 in C08.  As in Sec. 3.1, we re-scale the observed density measure to compare directly with our own density measure, M$_{\mathrm{sph}}$.  The DEEP2 survey does not probe clusters, but includes somewhat larger groups at $z$$\sim$1 than the GOODS fields (Gerke et al. 2005), so we align M$_{\mathrm{sph}}$ with the density measure in C08 such that the highest local density measurement in DEEP2 is in a DM sphere of 2$\times$10$^{13}$ M$_{\odot}$.  We again assume that there is a one-to-one relationship between the 2D ($\Sigma_{gal}$ in C08) and 3D (M$_{\mathrm{sph}}$ in this work)  local density measures.

\begin{figure}
\includegraphics[trim= 0mm 0mm 0mm 0mm, clip]{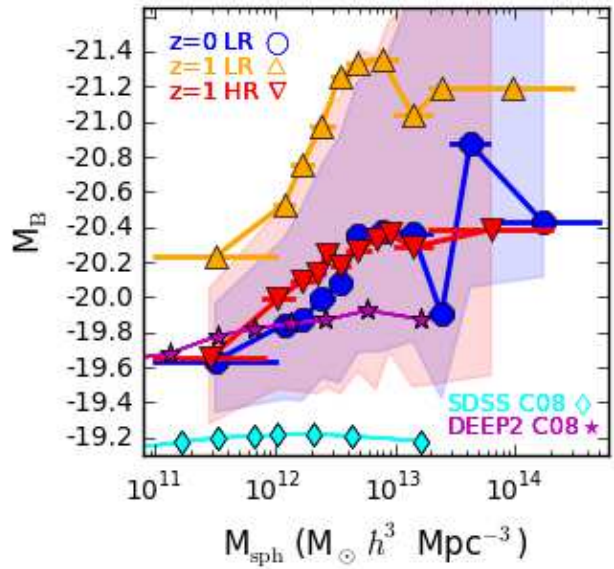}\\%
\caption{The median values of M$_{\mathrm{B}}$ of blue galaxies in M$_{\mathrm{sph}}$ bins--our measure of local density.  As in Figure \ref{fig-Cooper}, we only plot galaxies with 5$\times$10$^{9}$ $<$ M$_*$/M$_{\odot}$, but do not use a minimum brightness cut and instead use a color cut:  $g-r$ $<$ 0.64 at $z$$=$0 and $g-r$ $<$ 0.57 at $z$$=$1.  The M$_{\mathrm{B}}$ of galaxies in each M$_{\mathrm{sph}}$ bin are similar at $z$=0 and $z$=1 HR, and decrease with increasing local density. }\label{fig-cmass}
\end{figure}

We find that sSFR decreases with increasing local density (M$_{\mathrm{sph}}$) at both $z$=0 and $z$=1 (top panel of Figure \ref{fig-Cooper}).  The $z$=0 sSFR-density relation is in very good agreement with the C08 SDSS-B sample's relation.  The shape of both the $z$$=$1 LR and HR sSFR-density relations match that of the DEEP2-C sample, although the absolute value of the sSFR is low by a factor of 4 (LR) to 8 (HR).  In order to quantitatively match the observed sSFR in C08 (and SFR), we would need to raise our SFRs, which would probably require changing some of the variables in our star formation prescription.  While it is difficult to predict how changing our thresholds for star formation or our star formation efficiency would affect our simulation at low redshift, it is possible that, for example, lowering the star formation efficiency could delay star formation in our simulation, resulting in lower-mass galaxies and more gas available for higher SFRs at later times.  Testing a series of these values in our zoom-in boxes is beyond the scope of this work, but an array of lower-resolution cosmological simulations with these types of variations have been compared using a smoothed-particle hydrodynamics (SPH) code in Schaye et al. (2010).  Changing our star formation efficiency would similarly affect halos of all masses and in all environments, so we expect the redshift dependence of the SFR-density relation to be robust to changes in the star formation efficiency.

In the bottom panel of Figure \ref{fig-Cooper}, at $z$=0, our SFR-density relation is nearly flat until M$_{\mathrm{sph}}$ $>$ 1.3$\times$10$^{13}$ M$_{\odot}$, at which point the SFR drops steeply with M$_{\mathrm{sph}}$.  While the C08 result (SDSS-B sample) drops by a similar fraction, it does so more smoothly and gradually.  At $z$$=$1, our SFR-density relation rises and falls more than in C08, but at M$_{\mathrm{sph}}$$\ge$2$\times$10$^{12}$ M$_{\odot}$, the shape of our SFR-density relation is in decent agreement with C08.  As with the $<$SFR$>$-density relation (Figure \ref{fig-elbaz}), we find that at $z$=1 the 75th percentile of the SFR of galaxies initially increases dramatically (by a factor of more than 5) as M$_{\mathrm{sph}}$ increases to 8$\times$10$^{12}$ M$_{\odot}$, while in this figure the 25th percentile of SFR stays flat.  In contrast, at $z$$=$0 the 75th percentile of the SFR of galaxies increases by less than a factor of 3.  This indicates that a population of strongly-star-forming galaxies is driving the increasing SFR-density relation at $z$=1. 

C08 find that their observed reversal of the SFR-density relation is largely due to the fact that the mean M$_{\mathrm{B}}$ of the blue galaxies in their samples decreases with local density at $z$$=$1 but is flat with local density at $z$=0.  They find that M$_{\mathrm{B}}$ is correlated with M$_*$, with scatter, so we can loosely rewrite their result to say that the M$_*$ of the blue galaxies in their sample increases with local density at $z$$=$1 but is flat at $z$$=$0.  Blue galaxies dominate their $z$$=$1 sample, so the increase in SFR with increasing local density reflects the M$_{\mathrm{B}}$-density relation (or M$_*$-density relation).  C08 claim that at $z$=1 only their highest local density bin has enough red galaxies to flatten the SFR- and M$_{\mathrm{B}}$-density relations.  At $z$$=$0, blue galaxies do not completely dominate the sample, so an increasing fraction of red galaxies will cause the median SFR to decrease with local density.  Also, M$_{\mathrm{B}}$ does not vary strongly with local density, so there is no strongly star-forming population driving an increase in the median SFR with increasing local density.   

\begin{figure*}
\begin{center}
\includegraphics[trim= 0mm 0mm 0mm 0mm, clip]{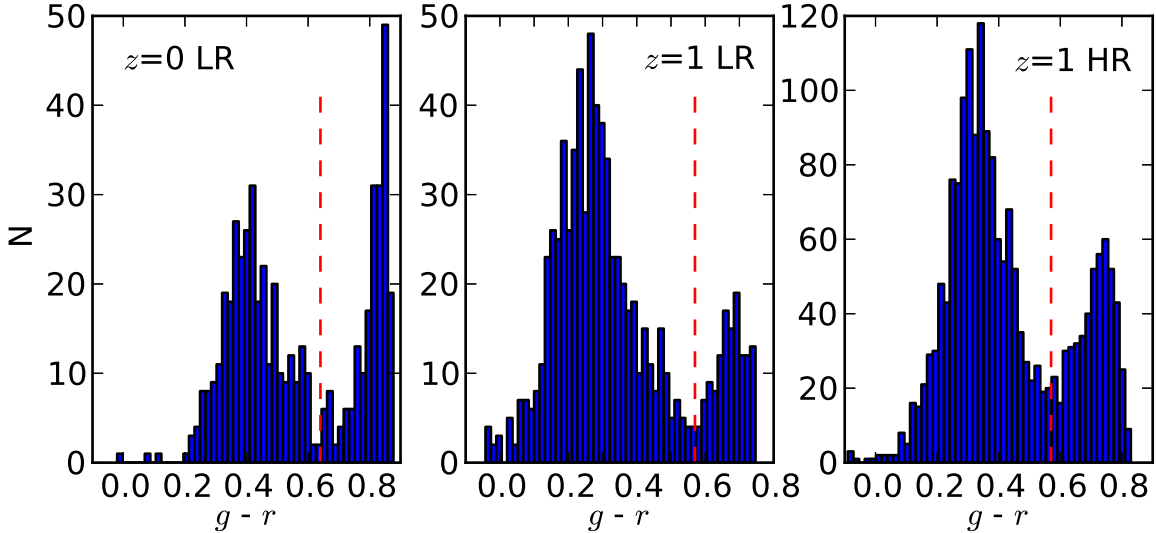}\\%
\caption{The $g$ $-$ $r$ color distribution of galaxies at the $z$$=$0 LR output, the $z$$=$1LR output, and the $z$$=$1 HR output (from left to right).  The panels show all galaxies with 5$\times$10$^{9}$ $<$ M$_*$/M$_{\odot}$.  The bimodal distribution is visible in all panels.  The red dashed lines denote the $g$ $-$ $r$ color we have chosen to split the blue (star-forming) from the red (passive) galaxies:  0.64 at $z$$=$0 and 0.57 at $z$$=$1. }\label{fig-gmrcolor}
\end{center}
\end{figure*}

In Figure \ref{fig-cmass} we plot the median M$_{\mathrm{B}}$ versus M$_{\mathrm{sph}}$ for blue galaxies ($g-r$ $<$ 0.64 at $z$$=$0 and $g-r$ $<$ 0.57 at $z$$=$1, see Figure \ref{fig-gmrcolor}) with 5$\times$10$^{9}$ $<$ M$_*$/M$_{\odot}$ (no minimum brightness cut).  There are several points to discuss in comparing our M$_{\mathrm{B}}$-density relation to that in C08 (their Figure 17).  First, in our simulations M$_{\mathrm{B}}$ decreases (and M$_*$ increases) with local density (M$_{\mathrm{sph}}$) at both redshifts.  In the $z$$=$1 HR sample the median M$_{\mathrm{B}}$ of galaxies in the lowest M$_{\mathrm{sph}}$ bin agrees with that in C08, but quickly moves to brighter M$_{\mathrm{B}}$ values as the local density increases.  The $z$$=$1 LR blue sample always has brighter median M$_{\mathrm{B}}$ values than either the $z$$=$1 HR or the C08 galaxies, and M$_{\mathrm{B}}$ increases the most steeply in the $z$$=$1 LR blue sample.  At $z$$=$0 the median M$_{\mathrm{B}}$ of our simulated galaxies is always brighter than that of the C08 SDSS sample, and decreases with increasing local density.  As we do not include dust in our simulations, we would expect our galaxies to be somewhat brighter than those in C08 (somewhere between 0.5-1.2 dex; Shao et al. 2007).  Also, the same adjustment to our simulation that we have proposed earlier in this Section may bring our results into closer agreement with those of C08--decreasing star formation efficiency will increase the M$_{\mathrm{B}}$ of galaxies.  

\subsection{The color-density relation}\label{sec:Q12}

We now find the red fraction of the simulated galaxies as a function of local density (M$_{\mathrm{sph}}$) at $z$=0 and $z$=1.  In Figure \ref{fig-gmrcolor} we plot the $g$ $-$ $r$ color distribution of the galaxies in our three outputs (5$\times$10$^{9}$ $<$ M$_*$/M$_{\odot}$).  There is a bimodal distribution in all of the galaxy samples, and we have chosen to split the blue (star-forming) and red (quiescent) populations at $g$ $-$ $r$ $=$ 0.64 at $z$$=$0 and $g$ $-$ $r$ $=$ 0.57 at $z$$=$1.  As our simulation does not have dust, this single color-cut is sufficient to differentiate star-forming from quiescent galaxies.  We tested that varying the value of the color-cut by 10\% has little quantitative impact on our results.  

\begin{figure}
\includegraphics[trim= 0mm 7mm 0mm 0mm, clip]{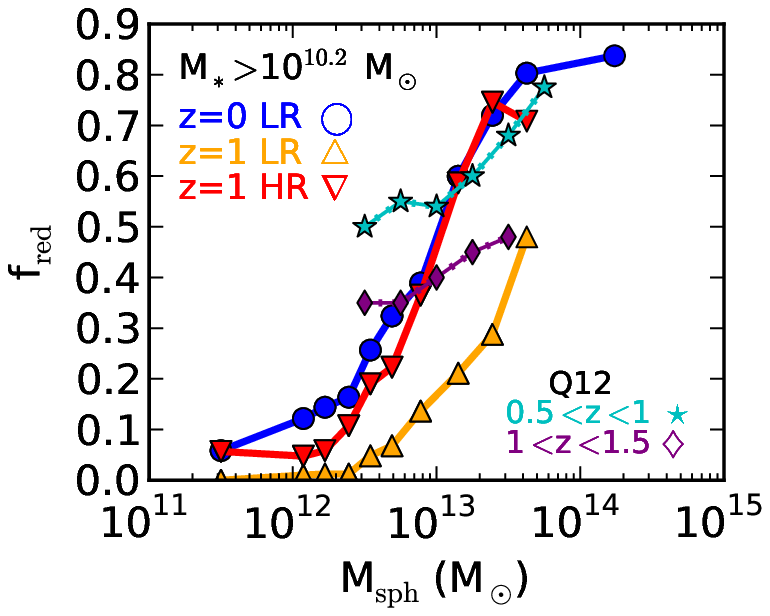}\\%
\includegraphics[trim= 0mm 0mm 0mm 0mm, clip]{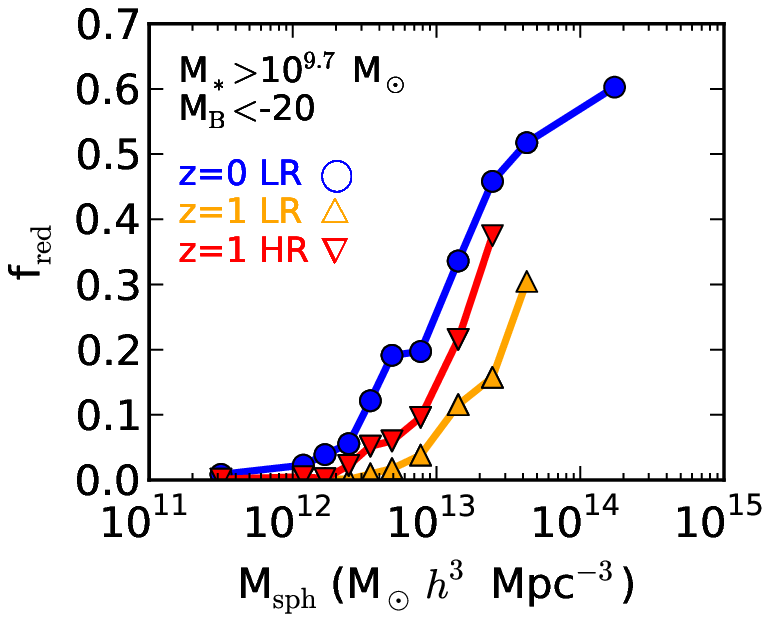}%
\caption{The fraction of red (quiescent) galaxies in bins of M$_{\mathrm{sph}}$ (local density).  \textbf{Top:}  We use the same low-mass cut as in Q12, so include all galaxies with 10$^{10.2}$ $<$ M$_*$/M$_{\odot}$.  We overplot the results from the Q12 paper that bracket $z$=1.  \textbf{Bottom:}  Using the same galaxy sample as in Figure \ref{fig-elbaz}.  The red fraction is much lower in this M$_{\mathrm{B}}$ $\le$ -20 sample, but still increases with increasing M$_{\mathrm{sph}}$ at both $z$=0 and $z$=1.  }\label{fig-Quadri}
\end{figure}

We plot the fraction of red (quiescent) galaxies in our samples in Figure \ref{fig-Quadri}.  At both $z$=1 and $z$=0, the fraction of red galaxies increases with increasing M$_{\mathrm{sph}}$.  Although there is little evolution in this figure from $z$$=$1 to $z$$=$0, in the $z$$=$1 LR sample, the fraction of red galaxies is always smaller than in the $z$$=$0 sample, and the $z$$=$1 HR sample does not reach the maximum red fraction found in the $z$$=$0 sample.  

We compare our results to those of Q12, who use a mass-selected sample (log(M$_*$/M$_{\odot}$) $>$ 10.2) from the UKIDSS Ultra-Deep Survey to plot the quiescent fraction (quiescent galaxies are defined using a color-color cut) of galaxies as a function of local density in different redshift ranges.  They find that even in their highest redshift bin (1.5$<$$z$$<$2.0), the fraction of quiescent galaxies increases with local density.  As shown in Figure \ref{fig-Quadri}, our results clearly qualitatively agree with those of Q12 (and many others; e.g. McCarthy et al. 2007; Kurk et al. 2009; Wilson et al. 2009; Chuter et al. 2011).  As with our previous comparisons to observations, the alignment of the observed galaxy density and M$_{\mathrm{sph}}$ is somewhat arbitrary.  

While our main conclusion is that the color-density relation exists at both $z$$=$0 and $z$$=$1, we now address the quantitative differences between our quiescent fraction and the observed quiescent fraction in Q12.  Our red fraction is too low in low local density (M$_{\mathrm{sph}}$) regions and too high in high local density (M$_{\mathrm{sph}}$) regions in comparison to Q12.  This is not an uncommon difference between simulations and observations, and has been seen, for example, in Tonnesen et al. (2007).  Some of this may be due to the local density measure used by Q12:  the distance to the 8th nearest neighbor.  They use photometric redshifts, which have large uncertainties, and interlopers may damp out strong trends in the data (see discussion in Q12).  Focusing on our simulation, we have no dust reddening our star-forming galaxies.  Although Q12 use a color-color selection in order to limit the number of dusty star-forming galaxy interlopers in their quiescent sample, their quiescent fraction may still be slightly high.  Also, we may have too much cold gas in galaxies in lower density regions due to the well-studied cooling flow problem (e.g. Fabian 1994; McNamara 2002; Croton et al. 2006), which may lead to low levels of star formation in galaxies that would otherwise have no star formation.  In high density environments, like the inner regions of clusters, cold gas may be used too quickly to form stars, and, as we have suggested above, decreasing the star formation efficiency may also reduce this discrepancy between our simulation and observations.

We find that changing our galaxy selection criteria has no qualitative effect on our results.  We changed the color-cut differentiating blue (star-forming) and red (passive) galaxies and the mass range of galaxies we included in our galaxy samples, and only included central galaxies in our sample.  Perhaps most importantly, in the bottom panel of Figure \ref{fig-Quadri} we plot the red fraction of galaxies that fulfilled the E07 criteria:  M$_{\mathrm{B}}$ $\le$ -20 and 5$\times$10$^9$ $<$ M$_*$/M$_{\odot}$ $<$ 10$^{12}$.  E07 also find that that the color-density relation still exists at $z$$=$1.

\section{Discussion}
\subsection{Physical Insight}\label{sec:story}

Our first main result is that we reproduce the reversal of the SFR-density relation from $z$=0 to $z$=1.  We conclude that this is because at $z$=1 the entire 25th-75th percentile range of SFRs at a given environment shifts upward with increasing local density (M$_{\mathrm{sph}}$), shown as the shaded region in Figures \ref{fig-elbaz} \& \ref{fig-Cooper}.  In $z$=1 HR there are relatively more highly star-forming galaxies (the 75th percentile) and relatively fewer galaxies with low SFRs (the 25th percentile bottom of the shaded region) up to a peak at M$_{\mathrm{sph}}$$\sim$10$^{13}$ M$_{\odot}$ in Figure \ref{fig-elbaz}.  While there is a population of highly star-forming galaxies in the highest density bin in our sample at $z$$=$0, there are not enough galaxies with high SFRs to drive a dramatic increase in the median SFR.  As we argue in \S \ref{sec:C08}, this agrees with the observational finding by Cooper et al. (2006) of a population of brighter blue galaxies in high local density environments at $z$$\sim$1.  

We also find that, using the same galaxy sample that shows an evolution in the SFR-density relation, there is an increasing fraction of red, quiescent galaxies as M$_{\mathrm{sph}}$ increases (bottom panel of Figure \ref{fig-Quadri}).  Until the red fraction increases to 50\% of the galaxies in the sample, much of the median SFR value will depend upon the distribution of the SFRs of the star-forming galaxies.  Indeed, we can see that the red fraction of galaxies that fulfill the E07 criteria (M$_{\mathrm{B}}$ $\le$ -20, 5$\times$10$^9$ $<$ M$_*$/M$_{\odot}$) \textit{never} reaches 50\% at $z$=1 (Figure \ref{fig-Quadri}).  Therefore, there is no tension between the observations of the median SFR increasing with local density and the observations of the fraction of quiescent galaxies increasing with local density--they are measuring two different aspects of the galaxy population.  This result has been pointed out observationally in E07.

\subsubsection{What drives the SFR-density relation?}

\begin{figure}
\includegraphics[trim= 0mm 0mm 0mm 0mm, clip]{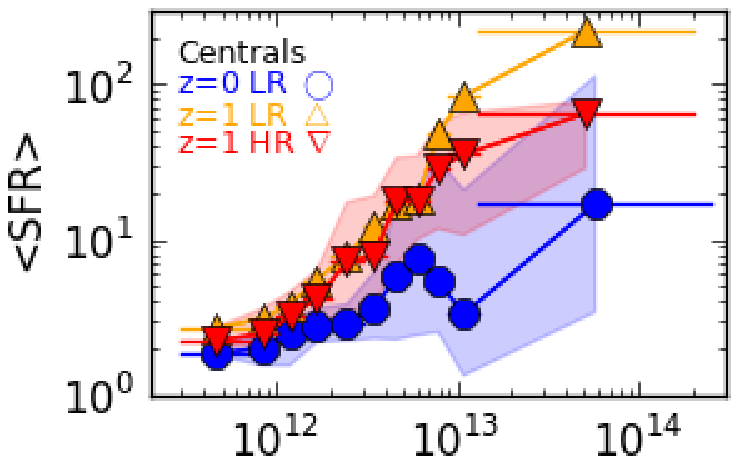}
\includegraphics[trim= 0mm 0mm 0mm 0mm, clip]{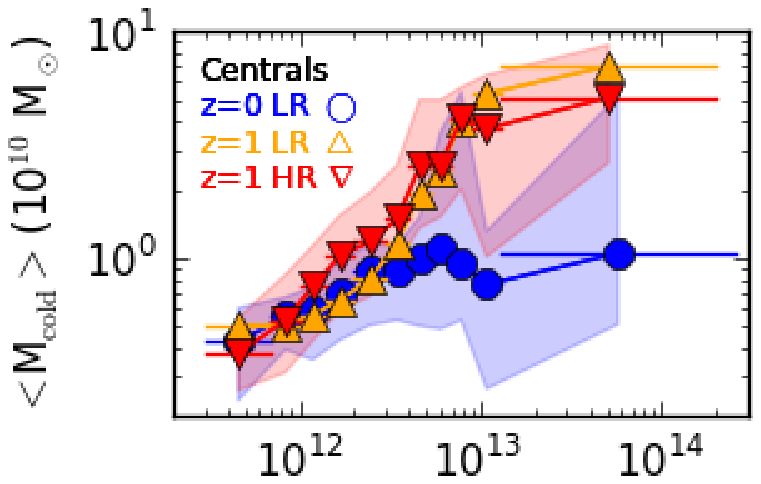}
\includegraphics[trim= 0mm 0mm 0mm 0mm, clip]{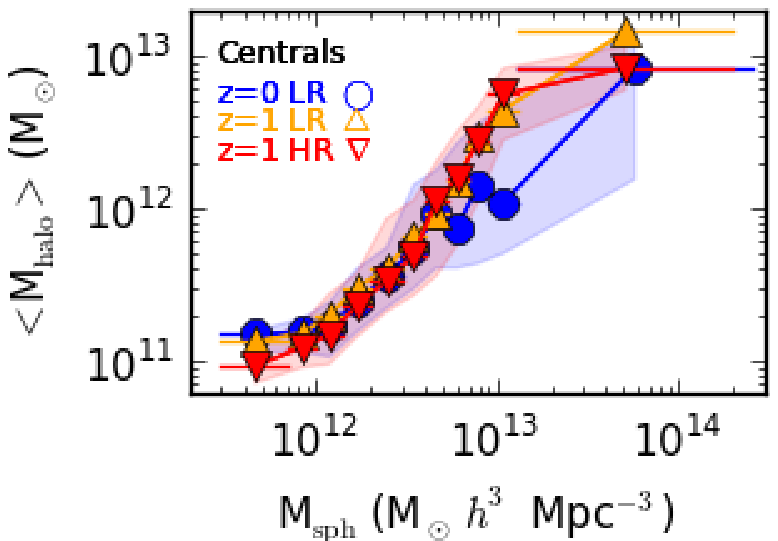}
\caption{The $<$SFR$>$, $<$M$_{\mathrm{cold}}$$>$ and $<$M$_{\mathrm{halo}}$$>$ of \textbf{central} galaxies with M$_{\mathrm{B}}$$\le$ -20 and 5$\times$10$^{9}$ $<$ M$_*$/M$_{\odot}$ $<$ 10$^{12}$ as a function of M$_{\mathrm{sph}}$.  Blue ($\bigcirc$) is low-resolution run at $z$=0, yellow ($\bigtriangleup$) is the low-resolution run at $z$=1, and red ($\nabla$) is the high-resolution run at $z$=1   \textbf{Top:}  Median (shaded region is 25th-75th percentiles) $<$SFR$>$ in each sphere binned by local density.  \textbf{Middle:}  Median (shaded region is 25th-75th percentiles) $<$M$_{\mathrm{cold}}$$>$ in each sphere binned by local density.  The galaxies with the highest SFR tend to have the highest M$_{\mathrm{cold}}$.   \textbf{Bottom:}  Median (shaded region is 25th-75th percentiles) of average mass in dark matter halos of central galaxies in each M$_{\mathrm{sph}}$ (dark matter sphere) bin. }\label{fig-dmhiall}
\end{figure}

We now look for a better understanding of why a much higher fraction of galaxies have high SFRs in higher density environments at $z$=1 than at $z$=0.  To do this we will discuss our findings in the context of a current standard theory of gas accretion and star formation.  First, we expect that a galaxy's cold gas reservoir is the fuel for star formation and therefore is an important factor in determining its SFR.  One current standard theory of gas accretion (e.g. Kere\v{s} et al. 2005; Dekel \& Birnboim 2006) contends that the amount of cold gas in a galaxy is determined by its dark matter halo mass, M$_{\mathrm{halo}}$.  Therefore, one would expect the M$_{\mathrm{halo}}$-M$_{\mathrm{sph}}$ relation to reflect the SFR-M$_{\mathrm{sph}}$ and M$_{\mathrm{cold}}$-M$_{\mathrm{sph}}$ relations.  

In order to compare our results with this theory we must focus only on central galaxies in their halos, as these are the only galaxies for which we could expect a relationship between gas and dark matter halo mass based on the two-mode theory of gas accretion.  A number of possible interactions can affect the gas, stellar, and dark matter mass of a satellite galaxy orbiting within a larger galaxy's halo, as discussed in Boselli \& Gavazzi (2006).  For satellite galaxies we find that the $<$SFR$>$ increases with density by a factor of 1.6 at $z$=1 LR, is flat with density at $z$=1 HR and decreases by a factor of 2 with density at $z$=0.  See Cen (2011c) for a detailed examination of why star formation ceases in satellite galaxies in this simulation.  

In the top panel of Figure \ref{fig-dmhiall}, we plot the median of the $<$SFR$>$ for only central galaxies in the simulation, which are the most massive galaxies in their halo.  These are galaxies that are beyond two virial radii from any more massive galaxy.  Most galaxies in our sample are central galaxies, but the fraction of satellites increases with M$_{\mathrm{sph}}$ (local density) until about half of the galaxies in the highest M$_{\mathrm{sph}}$ bin are satellites (27\% at $z$=0 LR, 68\% at $z$=1 LR and 53\% at $z$=1 HR).  Note that because of the large range in $<$SFR$>$, we plot the y-axis on a logarithmic scale.  When we focus only on central galaxies our results do not qualitatively change and we still see a dramatic change with redshift in the $<$SFR$>$-density relation.

In the middle panel of Figure \ref{fig-dmhiall} we examine the cold gas reservoir by plotting $<$M$_{\mathrm{cold}}$$>$ as a function of M$_{\mathrm{sph}}$.  This value measures the amount of gas that is, or will likely soon be, available for star formation.  For the low resolution outputs, this is the neutral HI gas that is within the r$_{200}$ of the parent halo, and for the high resolution output, this is the gas with T $<$ 10$^5$ K that is within 100 kpc of the galaxy (see Section \ref{sec:galaxyselection}).  As we expected, the shape of $<$M$_{\mathrm{cold}}$$>$ agrees well with that of the $<$SFR$>$ in all three outputs.

In the bottom panel of Figure \ref{fig-dmhiall} we plot the average dark matter halo mass ($<$M$_{\mathrm{halo}}$$>$) of galaxies at each M$_{\mathrm{sph}}$.  As defined in Section \ref{sec:galaxyselection}, the dark matter halo mass is the M$_{200}$ of the central halo--all the mass encompassed in a region with $\rho_{DM}/\rho_{crit} > $200.  A key point to remember is that M$_{\mathrm{sph}}$ is \textit{not usually} the mass of a single halo, it is the mass in a sphere with a radius of 1 comoving $h^{-1}$ Mpc (Section \ref{sec:localdensity}).  For reference, the M$_{200}$ of a single halo with r$_{200}$ = 1 $h^{-1}$ Mpc is $\sim$7$\times$10$^{13}$ $h^{-1}$ M$_{\odot}$.  

We first notice that at a specific local density (M$_{\mathrm{sph}}$), galaxies tend to reside in similar mass halos at $z$=1 and $z$=0.  For all three outputs, the general shape of the $<$M$_{\mathrm{halo}}$$>$ curve agrees with the general shape of the $<$SFR$>$ curve and the $<$M$_{\mathrm{cold}}$$>$ curve.  However, starting at M$_{\mathrm{sph}}$$\sim$4$\times$10$^{12}$ M$_{\odot}$, the difference between the halo masses at $z$=1 and $z$=0 is smaller than the difference in the $<$SFR$>$ and $<$M$_{\mathrm{cold}}$$>$.  

\begin{figure*}
\begin{center}
\includegraphics[trim= 0mm 0mm 0mm 0mm, clip]{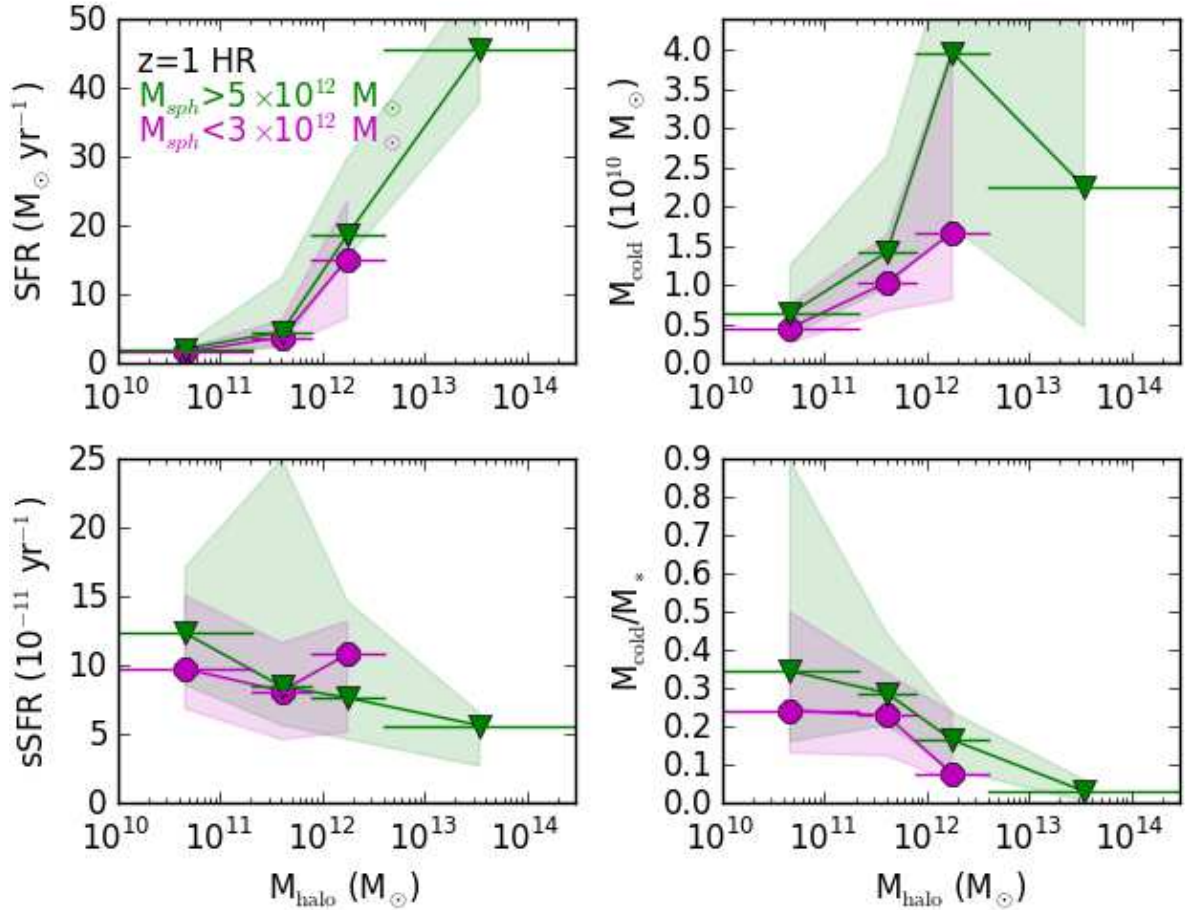}\\%
\end{center}
\caption{Galaxy properties as a function of M$_{\mathrm{halo}}$ at $z$=1 (HR) for central galaxies only.  We split the galaxy sample (M$_{\mathrm{B}}$ $<$ -20 and 5$\times$10$^{9}$ $<$ M$_*$/M$_{\odot}$) into low local density (M$_{\mathrm{sph}}$ $<$ 3$\times$10$^{12}$ M$_{\odot}$: purple circles) and high local density (M$_{\mathrm{sph}}$ $>$ 5$\times$10$^{12}$ M$_{\odot}$:  green triangles) samples.  In general, the SFR and M$_{\mathrm{cold}}$ change more strongly with M$_{\mathrm{halo}}$ than with local density (M$_{\mathrm{sph}}$).  However, galaxies at higher local densities may have slightly higher SFRs and M$_{\mathrm{cold}}$.}\label{fig-z1mhalo}
\end{figure*}

Interpreting the panels of Figure \ref{fig-dmhiall} within the cold-hot two mode theory of gas accretion, it seems that SFR depends on the amount of cold gas, which in turn depends strongly on the halo mass.  The cold gas mass increases from the lowest halo masses to higher halo masses because there is simply more gas in more massive halos, although M$_{\mathrm{cold}}$/M$_*$ decreases with increasing halo mass.  The fraction of cold gas with respect to the total amount of gas in the halo decreases with increasing halo mass, as found in Kere\v{s} et al. (2005), and we see that from the penultimate to the final bin $<$M$_{\mathrm{halo}}$$>$ increases by a higher fraction than either $<$SFR$>$ or $<$M$_{\mathrm{cold}}$$>$ in all three outputs.  Kere\v{s} et al. (2005) find that hot mode accretion dominates in halos above 10$^{11.4}$ M$_{\odot}$, the median halo mass of galaxies residing in environments with M$_{\mathrm{sph}}$$\sim$2.5$\times$10$^{12}$ M$_{\odot}$, the highest density as which we find agreement between the $z$=1 and $z$=0 outputs.  In this interpretation, the lower fraction of cold gas accretion could be reducing the cold gas available at $z$=0 with respect to $z$=1, therefore lowering SFRs at higher halo mass because of reduced supply.  

This interpretation could be tested by examining higher redshifts and determining how the gas supply and SFR evolve.  If Keres et al. (2005) are correct that at $z$=3 the cold gas fraction decreases with increasing halo mass (although the fraction is consistently higher than at lower redshift) our results suggest that the SFR-density relation should be steeper with increasing density and halo mass at higher redshifts.  This would be a good test of whether cold accretion indeed decreases with halo mass at high redshift.

\subsubsection{A Closer Examination of Halo Mass}

\begin{figure*}
\begin{center}
\includegraphics[trim= 0mm 0mm 0mm 0mm, clip]{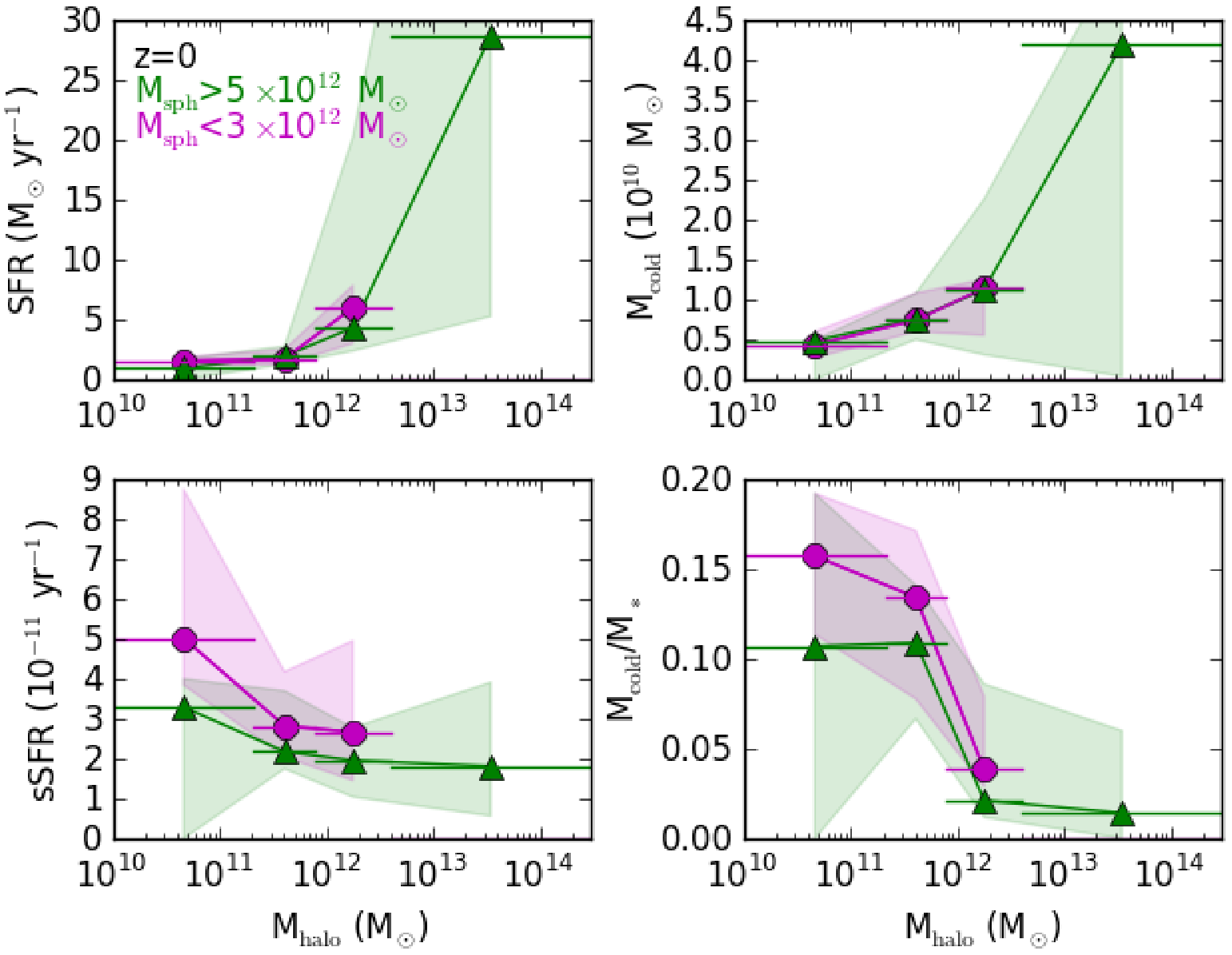}\\%
\end{center}
\caption{As Figure \ref{fig-z1mhalo}, but for galaxies at $z$=0.  The differences between galaxies at low- and high-densities are more pronounced at this lower redshift, and indicate that SFR and M$_{\mathrm{cold}}$ are both lower in galaxies in higher local density regions.}\label{fig-z0mhalo}
\end{figure*}

In the above section, we have shown that we can fit our data into a theory in which SFR depends on M$_{\mathrm{halo}}$ and M$_{\mathrm{halo}}$ depends on M$_{\mathrm{sph}}$.  However, this scenario does not explain the increasing sSFR with increasing M$_{\mathrm{sph}}$ show in Figure \ref{fig-essfr}.  Therefore in this section we will look directly at how the environment, M$_{\mathrm{sph}}$, affects galaxies with the same M$_{\mathrm{halo}}$.

To compare the importance of M$_{\mathrm{halo}}$ to M$_{\mathrm{sph}}$ in determining galaxy SFRs we split our central galaxy sample (central galaxies with M$_{\mathrm{B}}$ $\le$ -20 and 5$\times$10$^{9}$ $<$ M$_*$/M$_{\odot}$) into low local density (M$_{\mathrm{sph}}$ $<$ 3$\times$10$^{12}$ M$_{\odot}$) and high local density (M$_{\mathrm{sph}}$ $>$ 5$\times$10$^{12}$ M$_{\odot}$) subsamples.  This brackets the M$_{\mathrm{sph}}$ bin at which the $z$=1 and $z$=0 values diverge in Figure \ref{fig-elbaz}.  In Figures \ref{fig-z1mhalo} \& \ref{fig-z0mhalo} we then plot galaxy properties in these two sub-samples as a function of M$_{\mathrm{halo}}$ at $z$=1 (Figure \ref{fig-z1mhalo}:  we only plot the galaxies in the HR simulation for easier viewing, and find the same qualitative results for the LR output) and at $z$=0 (Figure \ref{fig-z0mhalo}).  It is useful to plot SFR and M$_{\mathrm{cold}}$ to compare the absolute values at different local densities, and plotting SFR and M$_{\mathrm{cold}}$ divided by M$_*$ allows us to normalize for the different M$_*$ distributions in the high and low density subsamples.  Clearly there will not be halos more massive than 3$\times$10$^{12}$ M$_{\odot}$ residing in regions with M$_{\mathrm{sph}}$ $<$ 3$\times$10$^{12}$ M$_{\odot}$, so the low density data set stops at that mass.  For completeness, we show the higher M$_{halo}$ galaxies in the high-density subset.  

We see that at $z$=1 (Figure \ref{fig-z1mhalo}), the low-density and high-density samples lie close together.  In particular, SFR increases by a factor of more than 5 as M$_{\mathrm{halo}}$ varies, but at a constant M$_{\mathrm{halo}}$ the difference between the low- and high-density samples is always less than a factor of 1.5.  The sSFR is very flat with M$_{\mathrm{halo}}$ such that the difference between the low- and high-density samples at a single M$_{\mathrm{halo}}$ can be greater than the difference between galaxies in different M$_{\mathrm{halo}}$ bins, particularly in the two highest M$_{\mathrm{halo}}$ bins.  However, the absolute value of the difference of the median sSFR in the high- and low-density samples is always small. The M$_{\mathrm{cold}}$ of the low- and high-density samples seems to be very different at high M$_{\mathrm{halo}}$, but some of this is probably due to the higher mass of galaxies at higher local densities, which is corrected for in the M$_{\mathrm{cold}}$/M$_{*}$ plot.  In addition to the median M$_{\mathrm{cold}}$ and M$_{\mathrm{cold}}$/M$_*$ tending to be higher at high M$_{\mathrm{sph}}$ than at low M$_{\mathrm{sph}}$, we see a tendency in all panels for the 25th-75th percentile range of the high-density galaxy sample to be higher than that of the low-density galaxy sample.  

At $z$=0 (Figure \ref{fig-z0mhalo}), the low- and high-density samples lie close enough to one another that again it is difficult to immediately determine whether the environment directly impacts the SFR and/or M$_{\mathrm{cold}}$ of galaxies.  In both the SFR and M$_{\mathrm{cold}}$ panels, the main difference between low- and high-density galaxies is the larger 25th-75th percentile range of galaxies in high-M$_{\mathrm{sph}}$ regions.  However, when comparing SFR to sSFR and M$_{\mathrm{cold}}$ to M$_{\mathrm{cold}}$/M$_*$, it becomes clear that the SFR and M$_{\mathrm{cold}}$ of high-M$_{\mathrm{sph}}$ galaxies is enhanced by a larger fraction of high-M$_*$ galaxies.  The differences between the low-density and high-density medians normalized by M$_*$ are larger than the absolute SFR and M$_{\mathrm{cold}}$.  Indeed, the difference between the sSFR of low- and high-density galaxy samples in the lowest or highest M$_{\mathrm{halo}}$ bin is similar to the difference between the sSFR in neighboring M$_{\mathrm{halo}}$ bins.  In addition, opposite to the trend at $z$$=$1, the 25th-75th percentile range of values differs in the sSFR plot with a consistently higher range of sSFRs in the low-density subsample than in the high-density subsample.  Also, when we normalize for M$_*$, we see that in the two lowest M$_{\mathrm{halo}}$ bins, the difference in median M$_{\mathrm{cold}}$/M$_*$ between galaxies in low- and high-density environments is twice that between galaxies in the same environment.  It seems that at $z$$=$0 local density has a similar level of impact on the SFR and M$_{\mathrm{cold}}$ of galaxies as the halo mass itself, particularly for lower mass halos (M$_{\mathrm{halo}}$$<$10$^{11.8}$ M$_{\odot}$).

When comparing Figures \ref{fig-z1mhalo} and \ref{fig-z0mhalo}, we see that the difference between low- and high-density galaxies in any individual M$_{\mathrm{halo}}$ bin tends to show the opposite trends at the two redshifts.  At $z$=0, galaxies in high local density regions tend to have lower SFR, sSFR, M$_{\mathrm{cold}}$ and M$_{\mathrm{cold}}$/M$_*$ than galaxies in low local density regions.  In contrast, at $z$=1, galaxies in high local density regions tend to have \textit{higher} sSFR and M$_{\mathrm{cold}}$/M$_*$ than galaxies in low local density regions.  At $z$=1 M$_{\mathrm{halo}}$ is a much stronger driver of galaxy SFR and M$_{\mathrm{cold}}$ than M$_{\mathrm{sph}}$, and we have provided clear evidence that the effect of the environment becomes more important at lower redshift.

\subsection{The Use of Hydrodynamical Simulations}\label{sec:simcomparison}

Why does our fully hydrodynamical simulation succeed in reproducing the observed increase in $<$SFR$>$ with increasing local density at $z$=1 while SAMs do not (as shown in Elbaz et al. 2007)?  Before we consider how the inclusion of hydrodynamics will affect our results, we consider the difference in dark matter resolution in our simulation versus the Millennium simulation.  In our low resolution refined region, dark matter particles are $\sim$10$^8$ M$_{\odot}$, about an order of magnitude better than the mass resolution of the Millennium simulation.  Thus, we can resolve smaller dark matter halos.  As we show in the bottom panel of Figure \ref{fig-dmhiall}, the increase in the median galaxy halo mass is one of the main causes of the increasing SFR with increasing M$_{\mathrm{sph}}$ at $z$$=$1.  In the lowest M$_{\mathrm{sph}}$, while we find $<$M$_{\mathrm{halo}}$$>$$\sim$1-2$\times$10$^{11}$ M$_{\odot}$, the number of $\sim$10$^{11}$ M$_{\odot}$ halos in the Millennium simulation may be significantly under-estimated due to mass resolution.  Guo et al. (2011) compare the dark matter halo mass functions in the Millennium simulation and in MS-II, which has more than two orders of magnitude better resolution.  They find that at M$_{\mathrm{halo}}$= 10$^{11}$ the mass density of halos in MS is 75\% that in MS-II (while at M$_{\mathrm{halo}}$= 10$^{12}$ they nearly agree).  If, due to this resolution-dependent underdensity of low-mass halos, $<$M$_{\mathrm{halo}}$$>$ is forced to be an order of magnitude higher at low local densities, there may be no increase in $<$SFR$>$ with increasing local density.  It may be possible to check whether resolution is the main issue by determining if SAMs can reproduce the reversal of the SFR-density relation when halos from the Millenium-II simulation are included in the analysis.  

Our different results may also be due to differences in the treatment of gasphysics, including SNe feedback processes
and large-scale structure collapse induced shock heating.  We find some correlation between M$_{\mathrm{sph}}$ and the M$_{\mathrm{cold}}$ and SFR of galaxies (Figures \ref{fig-dmhiall}, \ref{fig-z1mhalo}, \& \ref{fig-z0mhalo}).  Cen (2011c) discusses how gravitational heating can be important outside of virialized regions, for example in collapsed filaments and pancakes.  This may be more likely to affect the reservoir of cold gas available to galaxies in high-density large-scale environments.  Because we self-consistently include gravitational heating, our simulations include these affects while SAMS cannot.  

\subsection{AGN Feedback and Resolution}\label{sec:res}

Here we will address the effects of AGN feedback and resolution on our results.  First, as discussed in Section \ref{sec:boxes}, we do not include AGN feedback in this analysis.  Cen (2011c) included a feedback prescription in which star formation is suppressed by a factor $f$ = $1/(1 + (M_h/10^{13} M_{\odot})^{2/3})$ post simulation, which results in better agreement between the simulated and observed $z$=0 $r$-band luminosity functions.  This suppression factor only has a large effect on galaxies in halos with masses at or above 10$^{13}$ M$_{\odot}$, so as is clear from Figure \ref{fig-dmhiall}, it only has a strong effect on the highest density bin.  This AGN feedback implementation lowers the $<$SFR$>$ in Figure \ref{fig-elbaz} by 45\% in the final bin of $z$=1 LR, resulting in a flat $<$SFR$>$-density relation across the two highest density bins.  The change is 30\% in final bins of the $z$=1 HR and $z$=0 outputs, and the shape of the $<$SFR$>$-density relation stays unchanged.  The general shape of Figure \ref{fig-Cooper} also remains unchanged, although the decrease of SFR and sSFR is slightly steeper in the final two bins for all outputs, which is at densities higher than observed in C08.  The differences in all of our other figures when we include this AGN feedback are also minor with no qualitative change in our results.  

While it is useful to point out that our results do not change when we include this form of AGN feedback, because of uncertainties in how AGN feedback should be implemented it is more clear to present our results without post-processing. Importantly, it is not clear that using the same relation between AGN feedback and halo mass is appropriate at $z$=0 and $z$=1 (see discussion in the review by Kormendy \& Ho 2013 \& references therein).  If black holes grow before bulges, then the AGN might be more massive at $z$=1 than at $z$=0.  Kormendy \& Ho (2013) argue that black holes are twice as massive relative to the bulge mass using $z$$\ge$2 quasars (but see, e.g., Schulze \& Wisotzki 2014; Kisaka \& Kojima 2010).  If the suppression of SF by AGN feedback is directly proportional to AGN mass, then we can test the importance of this difference by suppressing SF by an extra factor of two at $z$=1.  Even if we suppress star formation in the $z$=1 outputs by a factor of two more than at $z$=0, the difference between $z$=1 and $z$=0 remains.  Testing every permutation of star formation suppression by AGN feedback is beyond the scope of this paper, but the fact that our results remain unchanged when including these two possible forms of AGN feedback lead us to believe that the large difference in the SFR-density relation we find at $z$=1 and $z$=0 is robust to variations in the prescription used to implement AGN feedback.  

Resolution can have several impacts on our results, and we are able to examine these effects by comparing our results in the $z$=1 LR output to the $z$=1 HR output.  Overmerging, discussed in detail in reference to these simulations in Lackner et al. (2012), is a resolution issue in which a lower resolution is in general more conducive to merging among galaxies in crowded environments, such as clusters of galaxies (White 1976; Moore et al. 1996).  Because overmerging will have more of an effect at larger local densities, the higher resolution run may naturally have a flatter M$_*$-density relation.  As we see in the bottom panel of Figure \ref{fig-essfr}, overmerging does not seem to have caused a difference between the low resolution and high resolution runs by $z$=1.  If overmerging occurs between $z$=1 and $z$=0, and results in galaxies that are too massive at $z$=0, the simulated SFRs will be higher than observed SFRs-which is indeed what we see in our comparisons.  However, lowering the SFRs in our simulation at $z$=0 will only serve to increase the difference in the relation between the two redshifts, so will only strengthen our result.  

Resolution may also impact the effect of SN feedback on galaxies.  In our implementation of supernovae feedback, energy is channeled into the 27 gas cells surrounding the star particle.  With lower resolution, this energy input will not heat the gas to as high temperatures and may allow it to cool relatively more quickly, hence may result in a reduced negative feedback effect in comparison to the higher resolution simulation.  As with overmerging, we see that this cannot be a large effect up to $z$=1 because the low and high resolution runs have similar galaxy mass distributions.  However, at $z$=1 the SFRs, particularly at high densities and in higher mass halos, are larger in the low resolution run than in the high resolution run.  By this late in the simulation the galaxies may be massive enough that in the high density central regions of halos rapid cooling is occurring.  While in the high resolution run supernovae feedback may result in lower SFRs, the lower resolution run has higher SFRs.  It is noteworthy that the difference in the low and high resolution runs at $z$=1 begins at higher M$_{\mathrm{sph}}$ than the difference between the $z$=1 outputs and the $z$=0 output.  \textit{These differences due to resolution do not outweigh the differences in the SFR-density relation at low and high redshift.}

It is not immediately clear how a difference in supernovae heating would affect $z$=0 SFRs.  If more gas is able to cool in the low resolution run and supernovae continue to heat the gas more efficiently in the high resolution run, the SFRs may be higher than in a high resolution run.  However, if more of the available gas is used quickly in the low resolution run, the SFRs may be lower at $z$=0 in the lower resolution run because of a lack of fuel.  Schaye et al. (2010) find a lower SFR at late times ($z$$\le$1) in higher resolution runs, in agreement with our findings.  As they use a lower resolution SPH simulation and a different feedback scheme, we cannot simply assume their results apply to our simulation, but as we have noted above, a lower SFR at $z$=0 better matches observations and strengthens our results.

\section{Conclusion}\label{sec:conclusion}

In this paper we examine the SFR-density and color-density relations at $z$=0 and $z$=1 in high resolution cosmological 
simulations with a refined region of size 21$\times$24$\times$20 $h^{-3}$ Mpc$^3$ centered on a cluster of mass $\sim 3\times 10^{14}$M$_{\odot}$ at $z$=0.  This refined region has a large range of local densities, which allows us to separate the halo-scale from the larger-scale environment.  We have utilized the high resolution (0.114-0.46 $h^{-1}$ kpc) in these simulations to examine the SFR of a large sample of galaxies in detail to determine how galaxy properties are related to environment on a 1 $h^{-1}$ Mpc scale at $z$=0 versus $z$=1.  Our results are summarized below.\\

\noindent 1)  At z=1, our simulations produce SFRs that increase strongly with increasing local density, as in the observations of E07 (Figure \ref{fig-elbaz}).  We also find strong evolution in our SFR-density relation from $z$=1 to a much flatter SFR-density relation at $z$=0.  In addition, we reproduce the results of Cooper et al. (2008) that the sSFR decreases with increasing density at both $z$=1 and $z$=0 (Figure \ref{fig-Cooper}).  However, as in Elbaz et al. (2007), when we only consider the massive galaxies in our sample, 5$\times$10$^{10}$ $<$ M$_*$/M$_{\odot}$, both sSFR and M$_*$ increase with local density (M$_{\mathrm{sph}}$) up to $\sim$3$\times$10$^{12}$ (Figure \ref{fig-essfr}).  We find that the increase in median SFR with increasing local density (M$_{\mathrm{sph}}$) at $z$=1 is caused by an increasing fraction of highly-star forming galaxies at a given local density at higher local densities.  Massive (5$\times$10$^{10}$ $<$ M$_*$/M$_{\odot}$) star-forming galaxies drive the increasing SFR-density relation.  
This agrees with observations of a population of $z$$\sim$1 galaxies with high SFRs at high local density (Cooper et al. 2006; Cooper et al. 2008).\\

\noindent 2)  Using a galaxy sample matched to Quadri et al. (2012) or to Elbaz et al. (2007), we find that the red fraction increases as a function of local density (M$_{\mathrm{sph}}$) at both $z$=0 and $z$=1 (Figure \ref{fig-Quadri}).  There is no tension between the dramatic difference in the SFR-density relation at $z$=1 compared to $z$=0 and the consistency of the color-density relation. \\

\noindent 3)  We find that the relationship between median M$_{\mathrm{halo}}$ and local density (M$_{\mathrm{sph}}$) is an important cause of the redshift-dependent behavior of the SFR-density relation (Figure \ref{fig-dmhiall}).  The SFR and M$_{\mathrm{cold}}$ values at $z$=0 begin to diverge from those at $z$=1 at lower M$_{\mathrm{sph}}$, $\sim$4$\times$10$^{12}$, than the M$_{\mathrm{halo}}$ values diverge, at M$_{\mathrm{sph}}$$\sim$8$\times$10$^{12}$.  We also show that the local environment on scales of 1 $h^{-1}$ Mpc is more important at $z$$=$0 than at $z$=1, and affects galaxy SFRs as much as halo mass at $z$=0 (Figures \ref{fig-z1mhalo} \& \ref{fig-z0mhalo}).  Finally, we find indications that the role of environment reverses from $z$=0 to $z$=1:  at $z$=0 high-density environments depress galaxy SFRs, while at $z$=1 high-density environments may serve to raise SFRs.

\acknowledgements
We thank an anonymous referee for thoughtful and helpful comments that greatly improved the paper.  We thank Dr. David Elbaz for providing observational data to us.  Computing resources were in part provided by the NASA High-
End Computing (HEC) Program through the NASA Advanced
Supercomputing (NAS) Division at Ames Research Center.  This work is supported in part by grant NASA NNX11AI23G.

\end{document}